\documentclass[aps,prb,reprint,twocolumn,nofootinbib]{revtex4-1}

\usepackage{amssymb, amsmath}
\usepackage{subfig}
\usepackage{graphicx}
\usepackage{color}
\usepackage{lineno,hyperref}
\usepackage{listings}
\usepackage{ulem}
\usepackage{amsmath}

\hypersetup{
    bookmarks=true,         % show bookmarks bar?
    unicode=false,          % non-Latin characters in Acrobat‚Äôs bookmarks
    pdftoolbar=true,        % show Acrobat‚Äôs toolbar?
    pdfmenubar=true,        % show Acrobat‚Äôs menu?
    pdffitwindow=false,     % window fit to page when opened
    pdfstartview={FitH},    % fits the width of the page to the window
    pdfauthor={},     		% author
    colorlinks=true,       	% false: boxed links; true: colored links
    linkcolor=blue,         % color of internal links
    citecolor=red,        	% color of links to bibliography
}

\usepackage[warn]{mathtext}

\usepackage{amstext}
\usepackage{placeins}
\usepackage{epstopdf}
\usepackage{multirow}
\usepackage[english]{babel}
\usepackage{tabularx}
\usepackage{tikz,bm}

\usetikzlibrary{arrows,shapes.geometric,backgrounds,positioning,calc}
\usepackage{caption}
\begin{document}

% \begin{frontmatter}

%\title{Magnetic characterization of C$_2$H and C$_2$F}
\title{Role of direct exchange and Dzyaloshinskii-Moriya interactions \\ in magnetic properties of graphene derivatives: C$_2$F and C$_2$H}
\author{V.~V. Mazurenko$^{1}$, A.~N. Rudenko$^{1,2}$, S.~A. Nikolaev$^{1}$, D.~S. Medvedeva$^{1}$, A.~I. Lichtenstein$^{1,3}$, and M.~I. Katsnelson$^{1,2}$}
\affiliation{$^{1}$ Theoretical Physics and Applied Mathematics Department, Ural Federal University, Mira Str.19, 620002 Ekaterinburg, Russia \\
$^{2}$ \mbox{Institute for Molecules and Materials, Radboud University, Heijendaalseweg 135, 6525 AJ Nijmegen, The Netherlands} \\
$^{3}$ Institute for Theoretical Physics, University of Hamburg, Jungiusstrasse 9, 20355 Hamburg, Germany}

\date{\today}

\begin{abstract}
According to the Lieb's theorem the ferromagnetic interaction in graphene-based materials with bipartite lattice is a result of disbalance between the number of sites available for $p_z$ electrons in different sublattices. Here, we report on another mechanism of the ferromagnetism in functionalized graphene that is the direct exchange interaction between spin orbitals. By the example of the single-side semihydrogenated (C$_2$H) and semifluorinated (C$_2$F) graphene we show that such a coupling can partially or even fully compensate antiferromagnetic character of indirect exchange interactions reported earlier [Phys. Rev. B {\bf 88}, 081405(R) (2013)]. As a result, C$_2$H is found to be a two-dimensional material with the isotropic ferromagnetic interaction and negligibly small magnetic anisotropy, which prevents the formation of the long-range magnetic order at finite temperature in accordance with the Mermin-Wagner theorem. This gives a rare example of a system where direct exchange interactions play a crucial role in determining a magnetic structure. In turn, C$_2$F is found to be at the threshold of the antiferromagnetic-ferromagnetic instability, which in combination with the Dzyaloshinskii-Moriya interaction can lead to a skyrmion state.    
\end{abstract}

\maketitle
\date{\today}

% \end{frontmatter}
\maketitle

\section{Introduction}       
The search for magnetism in graphene-based materials is an attractive research field promising for spintronics applications \cite{Han,Roche}. According to theoretical predictions, $sp$-electron magnetic semiconductors might also have much higher Curie temperatures than the conventional ones \cite{Edward,Zhou}. Activities in this direction stimulate synthesis and magnetic measurements of graphene with vacancies and different types of adsorbates \cite{Asenjo, Hong,McCreary,Sepioni,Nair2012,Nair2013,Brihuega}. Despite considerable efforts, the available experimental data on magnetism in graphene-based systems is still limited. From the theoretical perspective numerous first-principles studies\cite{Helm, Boukhvalov, Yazyev, Sahin, Santos} allow to provide a microscopic picture on the electronic structure of magnetic graphene in its ground state. The treatment of excited states constitutes the next important step in the description of the systems in question. This requires the construction and solution of the model electronic or spin Hamiltonians, which have been only marginally addressed in the literature \cite{Sofo,Rudenko,Ulybyshev}. Considerable nonlocal Coulomb correlations typical to graphene\cite{Wehling} and spin-orbit coupling\cite{Fabian1,Fabian2} complicate the consideration significantly. Finally, practically important aspects such as the role of temperature and external fields in the evolution of magnetic states also remain unclear. 

Single-side semifluorinated graphene (C$_2$F) is of special interest because such a system was recently realized in the experiment.\cite{Kashtiban} It opens a way for verification and correction of the theoretical models for this material, whose magnetic properties represent a complex interplay between different physical mechanisms. According to the Lieb's theorem\cite{Lieb} formulated for bipartite lattice, C$_2$F should be ferromagnetic since the fluorine atoms adsorb at the same sublattice of carbon atoms. Such model, however, does not take into account considerable modification of the electronic structure upon fluorination. More reliable description of the magnetism in C$_2$F has been recently proposed using first-principles DFT calculations \cite{Rudenko} of the isotropic exchange interactions, predicting a frustrated ground state in C$_2$F. Such approach yet ignores many-body and relativistic effects, which might be crucial in the formation of magnetism.

In this paper we perform a systematic theoretical characterization of the C$_2$F and C$_2$H systems by constructing the low-energy models with spin-orbit coupling. In both cases the spin orbitals described by magnetic Wannier functions form a triangular lattice with short-range nearest-neighbour interactions in the case of C$_2$F and long-range interactions in the case of C$_2$H. The estimation of the on-site and inter-site Coulomb interactions suggest significant spatial charge correlations in both systems that are an order of magnitude larger than those in transition-metal compounds. Remarkably, we find a strong ferromagnetic {\it direct} interaction between the neighbouring spin orbitals. Due to a delicate balance between the direct ferromagnetic exchange interaction and kinetic Anderson's superexchange, our Hartree-Fock simulations show that C$_2$F can demonstrate both the 120$^{\circ}$ N\'eel and ferromagnetic states having comparable energies.  At the same time a robust ferromagnetic solution within the mean-field approximation is found for C$_2$H, contrary to previous predictions \cite{Rudenko}. The analysis of the anisotropic superexchange couplings reveals a strong Dzyaloshinskii-Moriya interaction (DMI) of about 1 meV between the nearest spins in C$_2$F. According to the Monte Carlo simulations, DMI can lead to a skyrmion state at finite temperatures and magnetic fields.

\begin{figure}[h!]
\includegraphics[width=0.47\textwidth,angle=0]{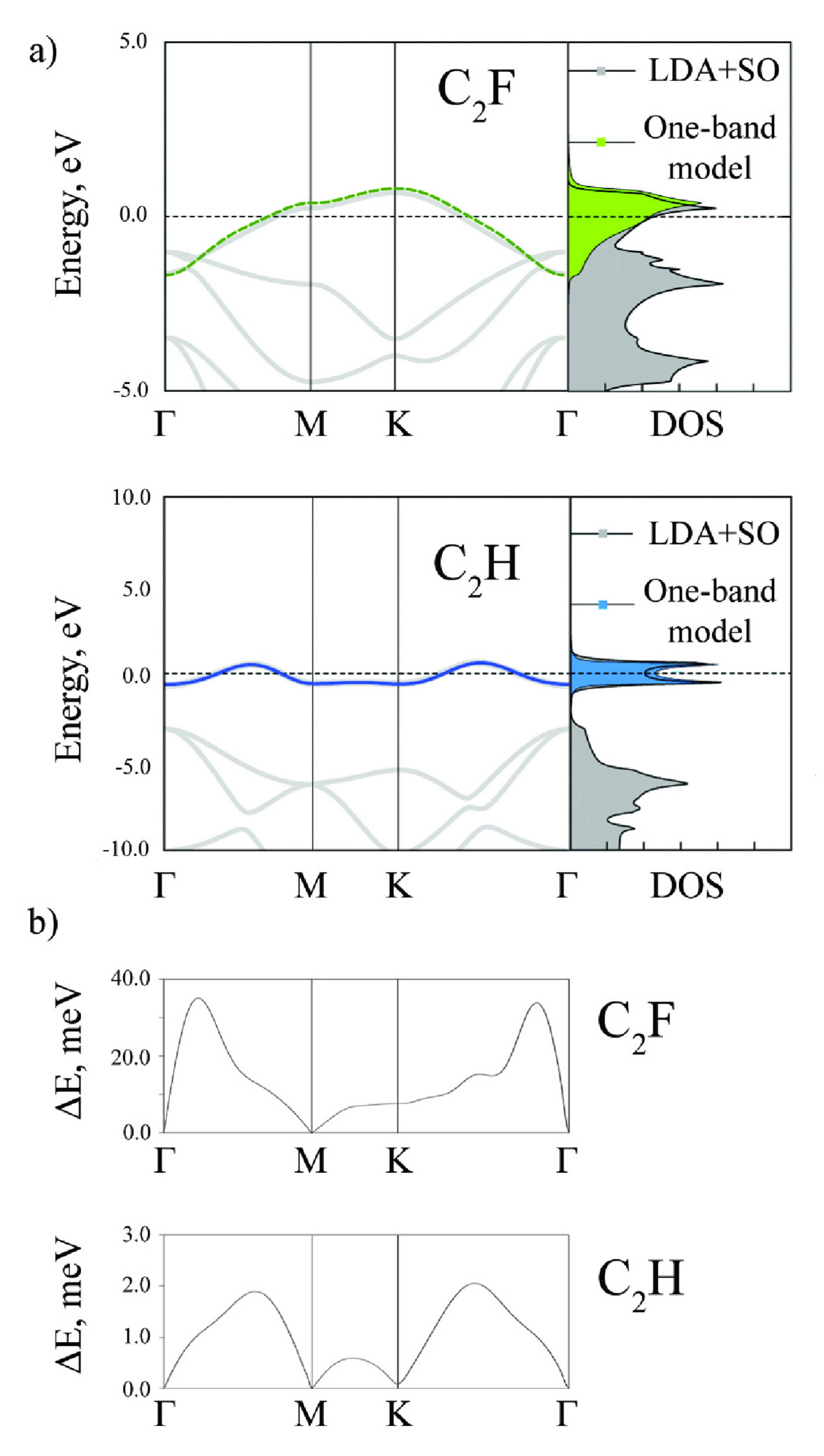}
\caption{(Color online) (a) Total densities of states and band structures of C$_{2}$F and C$_{2}$H calculated by using LDA+SO method and their comparison with the one-band model. (b) Band splitting $\Delta E$ due to spin-orbit coupling calculated for the one-band model.}
\label{LDA}
\end{figure}

\section{DFT results}
The electronic structure calculations were performed within the plane-wave pseudopotential method as implemented in the {\sc quantum-espresso} simulation package.\cite{Espresso}
Exchange and correlation effects were taken into account using the local density approximation (LDA).\cite{LDA} Spin-orbit (SO) coupling was included on the basis of fully relativistic pseudopotentials. We employed an energy cutoff of 50 Ry for the plane-wave basis and 400 Ry for the charge density, as well as a (64$\times$64) {\bf k}-point mesh. The surface layers were separated by a vacuum region of 40 \AA \, and fully relaxed. To construct the Hamiltonian in the (spinor) Wannier function basis we used the maximally localized Wannier function procedure\cite{Marzari} as implemented in the {\sc wannier90} package.\cite{wannier90}  

The calculated LDA+SO band structures and densities of states are presented in Fig.~\ref{LDA}(a). One can see that in the case of C$_2$H there are two well-separated bands at the Fermi level [with a small band splitting due to spin-orbit coupling, Fig.~\ref{LDA}(b)]. These bands demonstrate the maxima between $\Gamma-M$ and $\Gamma-K$ high-symmetry points. As we will show below, such a band behavior results in the long-range hopping integrals. However, it is not the case for the C$_2$F system, where the bands at the Fermi level slightly overlap with other bands at the $\Gamma$ point. Nevertheless, construction of the minimal low-energy model for the relevant bands at the Fermi level is confirmed by comparison with the many-orbital tight-binding model described in Ref.~\onlinecite{Rudenko}.

\begin{figure*}[t]
\includegraphics[width=1.0\textwidth,angle=0]{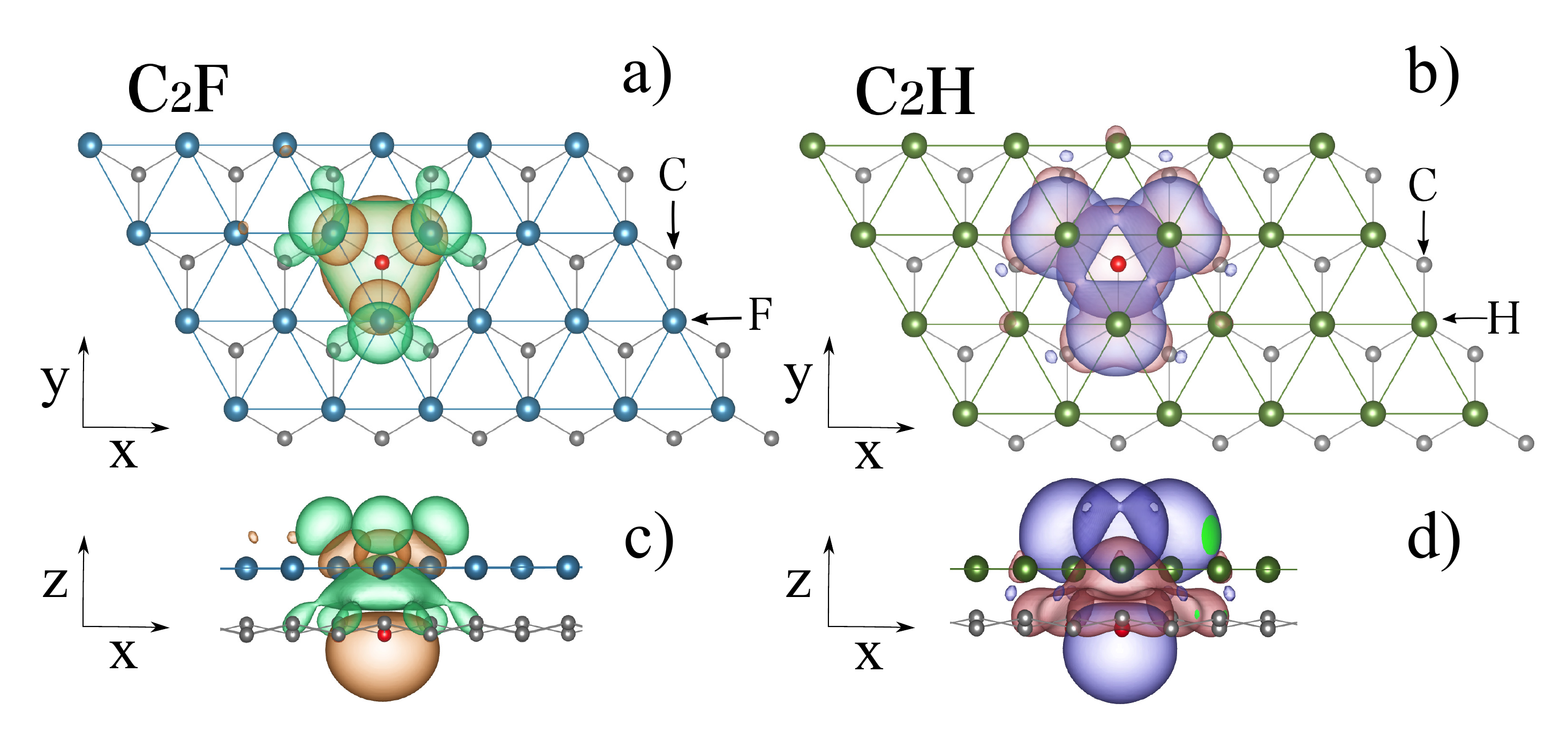} 
\caption{(Color online) Wannier functions describing the band at the Fermi level in C$_{2}$F (a,c) and C$_2$H (b,d). Red sphere denotes the center of the Wannier orbital.}
\label{ris:wannier_functions}
\end{figure*}

{\it Wannier functions.}
To parametrize the LDA+SO spectra we have constructed the maximally localized Wannier functions for the bands located close to the Fermi level. They are visualized in Fig.~\ref{ris:wannier_functions}. In agreement with the results of Ref.~\onlinecite{Rudenko} the Wannier functions are centered at the non-bonded carbon atoms. They are strongly delocalized in real space, that should be taken into account when analyzing the experimental data of the magnetic measurements.\cite{Form-Factor}  The corresponding band splitting due spin-orbit coupling for the one-band model is shown in Fig.~\ref{LDA}(b). As it is seen, it is an order of magnitude larger in the case of  C$_2$F.\cite{Fabian1,Fabian2}

\section{Low-energy model}

To describe electronic and magnetic properties of C$_{2}$H and C$_{2}$F we use the following tight-binding Hamiltonian taking into account spin-orbit coupling: 
\begin{eqnarray}
\hat{\cal H}=\sum_{i j,\sigma\sigma'}t_{ij}^{\sigma\sigma'}\hat{a}_{i \sigma}^{+}\hat{a}_{j \sigma'}+\frac{1}{2}\sum_{i ,\sigma\sigma'}U_{00} \,\hat{a}_{i \sigma}^{+}\hat{a}_{i \sigma'}^{+}\hat{a}^{}_{i \sigma'}\hat{a}^{}_{i \sigma} \nonumber \\ 
+ \frac{1}{2}\!\sum_{i j,\sigma\sigma'} U_{ij}\,\hat{a}_{i \sigma}^{+}\hat{a}_{j \sigma'}^{+}\hat{a}^{}_{j \sigma'}\hat{a}^{}_{i \sigma} + \frac{1}{2}\sum_{i j,\sigma\sigma'}J^{F}_{ij}\,\hat{a}_{i \sigma}^{+}\hat{a}_{j \sigma'}^{+}\hat{a}^{}_{i \sigma'}\hat{a}^{}_{j \sigma},
\label{Ham}
\end{eqnarray}
where $\hat{a}^{\dagger}_{i \sigma}$($a_{i \sigma}$) are the creation (annihilation) operators. $U_{00}$, $U_{ij}$ and $J^{F}_{ij}$ are local Coulomb, non-local Coulomb and non-local ($i \ne j$) exchange interactions, respectively. $t_{ij}^{\sigma \sigma'}$ is the element of the spin-resolved hopping matrix. The model Hamiltonian given by Eq.~(\ref{Ham}) can be solved by static (Hartree-Fock)\cite{Nikolaev} or dynamical (DMFT)\cite{DMFT} mean-field numerical techniques to reproduce experimentally observable spectra of electronic and magnetic excitations. This model is not only widely used in the simulations of physical properties of 3$d$ and 5$d$ metal compounds, but
can also be applied to the systems demonstrating $sp$-type magnetism.\cite{NaO2}

\begin{figure}[h!]
	\includegraphics[width=0.8\linewidth]{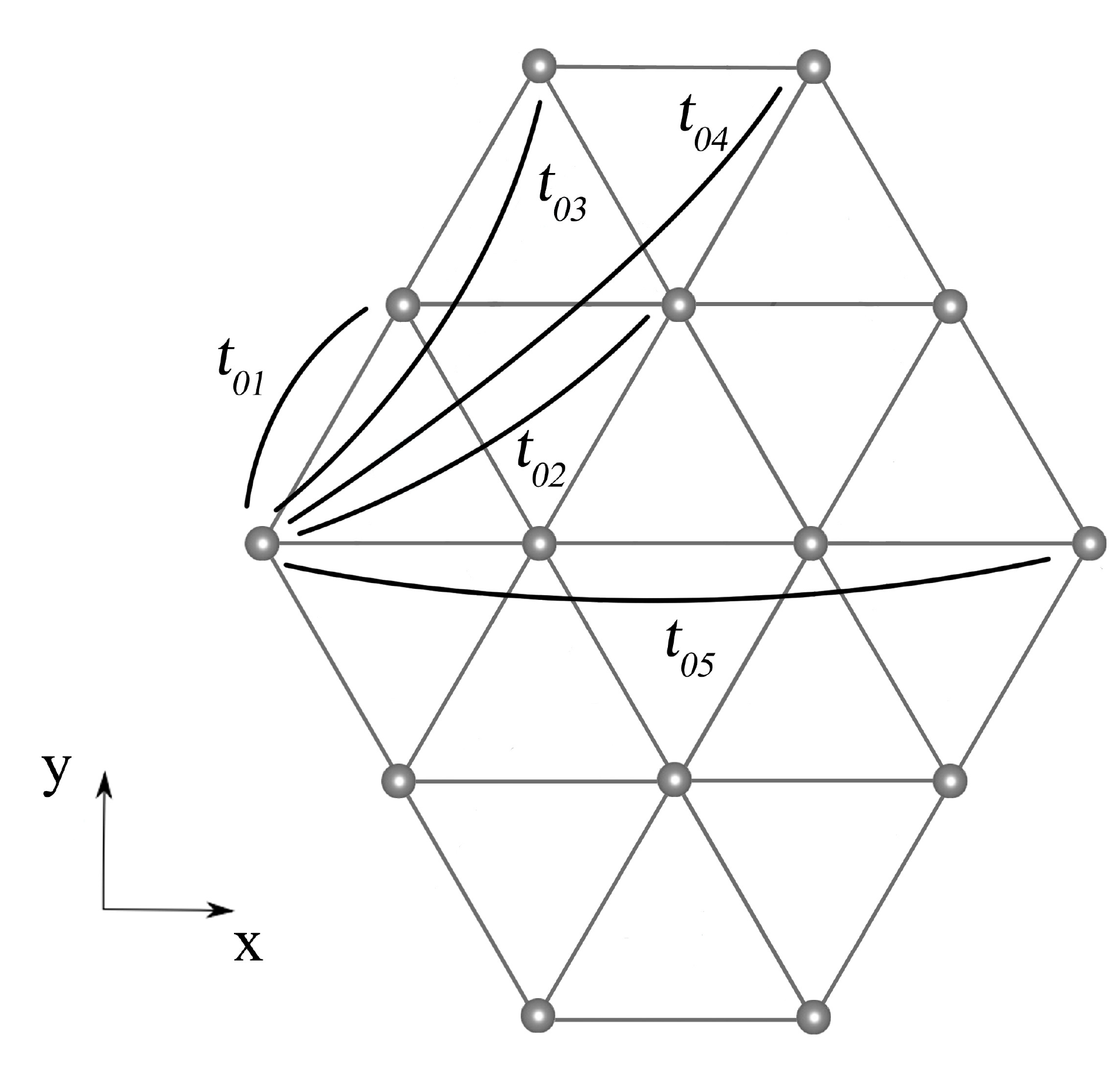} 
	\caption{Schematic representation of the hopping paths in the triangular model for C$_{2}$F and C$_{2}$H. The gray spheres denote the Wannier functions centered at non-bonded carbon atoms.}
	\label{lattice} 
	\end{figure}

The hopping matrix $t_{ij}^{\sigma \sigma'}$ was determined using a Wannier-parametrization of 
the first-principles LDA+SO Hamiltonian.
The obtained hopping parameters are presented in Table \ref{tab:hoppings} and schematically visualized in Fig.~\ref{lattice}.
One can see that in the case of the C$_2$F system the nearest-neighbor hopping ($t_{01}$) provides the dominant contribution, that results in the realization of an almost ideal triangular geometry for hopping integrals. For C$_2$H, the hopping matrix is less trivial with the largest parameters corresponding to the second- and third-nearest-neighbor interactions.
Such a difference in hopping integrals of the semifluorinated and semihydrogenated graphene is related to the 
various inter-atomic distances. The equilibrium carbon-carbon and carbon-adatom distances \cite{Rudenko} in C$_2$H are smaller by 4\% and 20\%, respectively, than those in C$_2$F. 

Importantly, there are non-diagonal and imaginary contributions to the hopping integrals between nearest neighbors. As one can see, such contributions are an order of magnitude larger in C$_2$F compared to C$_2$H. They originate from the spin-orbit coupling and, as we will show below, responsible for a non-zero Dzyaloshinskii-Moriya interaction. 

\begin{table*}
\caption{\label{tab:hoppings}Spinor representation of hopping integrals (in meV) calculated for C$_2$F and C$_2$H on the basis of the Wannier parametrization of the LDA+SO Hamiltonian.}
\begin{tabular}{ccc}
\hline
 & C$_{2}$F & C$_{2}$H\\
\hline
 & & \\
$t_{01}$ &
$ \begin{pmatrix} 
- 232.84 -  0.82i &	& 1.35 - 2.35i \\ 
- 1.35 - 2.35i & & -232.84 + 0.82i
\end{pmatrix}$ &
$\begin{pmatrix} 
38.98 + 0.02 	& & - 0.14 + 0.25i \\ 
0.14 + 0.25i 	& & 38.98 - 0.02i
\end{pmatrix} $ \\
 & & \\
\hline

 & & \\
$t_{02}$ &
$\begin{pmatrix} 
5.95 +  0i 	& & 0.65 - 0.37i \\ 
- 0.65 - 0.37i	& & 5.95 + 0i 
\end{pmatrix} $ & 
$ \begin{pmatrix} 
-114  + 0i &	& 0.04 - 0.02i \\ 
- 0.04 - 0.02i &	& -114 + 0i 
\end{pmatrix}$ \\
 & & \\
\hline

 & & \\
$t_{03}$ & 
$\begin{pmatrix} 
- 21.29 - 0.1i	& & 0.37 - 0.64i \\ 
- 0.37 - 0.64i	& & - 21.29 + 0.1i
\end{pmatrix} $ & 
$ \begin{pmatrix} 
- 98.05 + 0.03i &	& 0.01 - 0.01i \\ 
-0.01 - 0.01i	& 	& - 98.05 - 0.03i
\end{pmatrix}
$ \\
 & & \\
\hline

 & & \\
$t_{04}$ & 
$ \begin{pmatrix} 
- 10.70 + 0i &	& 0.39 - 0.31i \\ 
- 0.39 - 0.31i	& & - 10.70 + 0i
\end{pmatrix}$ & 
$\begin{pmatrix} 
27.92  + 0i &	& 0 + 0i\\ 
0  + 0i	&	& 27.92  +  0i 
\end{pmatrix}$ \\
 & & \\
\hline

 & & \\
$t_{05}$ & 
$\begin{pmatrix} 
- 10.40 + 0.04i & & 0.37  + 0i  \\ 
- 0.37  + 0i	& & - 10.40  -  0.04i
\end{pmatrix}$ & 
$\begin{pmatrix} 
11.86 + 0i 	& & -0.01  + 0i \\ 
0.01 + 0i 	& & 11.86  +  0i 
\end{pmatrix}$ \\
 & & \\
\hline
\end{tabular}
\end{table*}

The local ($U_{00}$) and non-local ($U_{ij}$) Coulomb interactions were determined in the static limit ($\omega=0$) using the constrained random-phase approximation (RPA) technique.\cite{cRPA1,cRPA2} Within this approach, the Coulomb interaction is screened by all the states except those described by the first term in Eq.~(\ref{Ham}). In the reciprocal-space representation the corresponding interaction reads
\begin{equation}
U({\bf q}) = [1-v({\bf q})P({\bf q)}]^{-1}v(\bf q),
\label{urpa}
\end{equation}
where $v(\bf q)$ is the Fourier-transform of the bare Coulomb interaction, which in 2D at ${\bf q}\rightarrow 0$ has the form\cite{DasSarma} $v({\bf q})=2\pi e^2/|{\bf q}|\kappa$ with $\kappa$ being the background (substrate) dielectric constant. At ${\bf q}\neq 0$, the bare interaction $v(\bf q)$ is evaluated between the Wannier functions using a standard expression for the Coulomb integrals (see, e.g., Ref.~\onlinecite{Mazurenko}). In Eq.~(\ref{urpa}), $P({\bf q})$ is the
static single-particle RPA polarizability calculated excluding transitions within the conduction band depicted in Fig.~\ref{ris:wannier_functions},
\begin{equation}
P({\bf q})=\frac{1}{\Omega}\sum_{i{\bf k}}^{\mathrm{occ}}\sum_{j{\bf k}'}^{\mathrm{unocc}}\frac{|\langle \Phi_{i{\bf k}}|e^{-i{\bf q}\cdot{\bf r}}|\Phi_{j{\bf k}'} \rangle|^2}{\varepsilon_{i{\bf k}}-\varepsilon_{j{\bf k}'} + i\eta },
\label{polariz}
\end{equation}
where ${\bf k}'={\bf k}+{\bf q}$ and the summation runs over the Brillouin zone involving transitions between the occupied and unoccupied states only. In Eq.~(\ref{polariz}), $\Omega$ is the volume of the unit cell, $i$ ($j$) denotes band indices, $\varepsilon_{i{\bf k}}$ ($\Phi_{i{\bf k}}$) is the eigenvalues (eigenvectors) of the full LDA Hamiltonian, and $\eta$ is a numerical smearing parameters chosen to be 10 meV. $P({\bf q})$ is evaluated on a ${\bf k}$-point mesh used in our LDA calculations. To estimate the non-local direct exchange integrals ($J^F_{01}$), we follow a slightly different procedure. In view of the relative smallness of $J^{F}_{01}$, the application of the constrained RPA scheme requires an extremely accurate Brillouin zone integration for the calculation of $P({\bf q})$ [Eq.~(\ref{polariz})], which cannot be achieved within the {\bf k}-point densities used in our paper. Instead, using the RPA procedure we can estimate the bare and fully screened $J^F_{01}$ with sufficient numerical accuracy, which are to be considered as upper and lower limits, respectively.

\begin{table}[!h]
\caption{The calculated local and non-local partially screened Coulomb interactions (in eV) for C$_2$F and C$_2$H. The two values of $J^{F}_{01}$ correspond to the fully screened and bare interactions.}
\label{tab:RPAresults}
\begin{tabular}{ccccc}
\hline
Interaction & C$_{2}$F & C$_{2}$H\\
\hline
$U_{00}$ &  5.16 &  4.69 \\
$U_{01}$ & 2.46 & 2.19  \\
$U_{02}$ & 1.66 & 1.11  \\
$U_{03}$ & 1.46 & 0.85 \\
$J^{F}_{01}$ (screened) & 0.018  &  0.034 \\
$J^{F}_{01}$ (bare) & 0.044    &  0.099 \\
%$J^{F}_{01}$ & 0.018  -- 0.044    &  0.034 -- 0.099 \\
\hline
\end{tabular}
\end{table}

The calculated Coulomb interactions are presented in Table \ref{tab:RPAresults}. In the case of C$_2$F both local and non-local couplings are slightly larger than those obtained for C$_2$H. It is related to the fact that the orbitals in
C$_2$F are more localized, resulting in a stronger repulsion. Indeed, linear spreads of the corresponding Wannier orbitals amounts to 1.76 \AA \, and 1.63 \AA \, for C$_2$H and C$_2$F, respectively. Importantly, there are strong long-range Coulomb interactions, which indicates significant spatial charge fluctuations in these graphene-based systems.
The direct exchange interaction between the nearest Wannier functions is much smaller than other Coulomb matrix elements. Nevertheless, as will be shown below, $J^{F}_{ij}$ plays a principal role in the formation of the magnetic states of C$_2$H and C$_2$F. 

\section{Magnetic interactions} 
Values of the calculated hopping integrals and Coulomb interactions correspond to the strong localization regime, $t_{ij} \ll U_{00}$, that allows us to construct a Heisenberg-type Hamiltonian for the localized spins $S=1/2$ within the superexchange theory developed by Anderson.\cite{Anderson} The corresponding spin model is given by
\begin{eqnarray}
\hat H_{spin} = \sum_{ij} J_{ij} \hat {\boldsymbol{S}}_{i} \hat {\boldsymbol{S}}_{j} + \sum_{ij} \boldsymbol{D}_{ij}  [\hat {\boldsymbol{S}}_{i} \times  \hat {\boldsymbol{S}}_{j}], 
\label{spinham}
\end{eqnarray}
where $\hat {\boldsymbol{S}}$ is the spin operator, $J_{ij}$ and $\boldsymbol{D}_{ij}$ are the isotropic and anisotropic (Dzyaloshinskii-Moriya) exchange  interactions. The summation over all pairs in Eq.~(\ref{spinham}) runs twice.
%Below we discuss the estimated isotropic and anisotropic exchange interactions. 

{\it Isotropic exchange interaction.}
In terms of the Hamiltonian parameters given by Eq.~(\ref{Ham}) the isotropic exchange interaction can be expressed in the following form \cite{Anderson, Aharony}
\begin{eqnarray}
J_{ij} = \frac{1}{\widetilde U} {\rm Tr_{\sigma}} \{ \hat t_{ji} \hat t_{ij} \} -  J^{F}_{ij},
\label{exch}
\end{eqnarray}
where the effective local Coulomb interaction is given as $\widetilde U = U_{00} - U_{ij}$.\cite{Mazurenko,Schuler} The first term is the antiferromagnetic Anderson's superexchange interaction, and %One should note that due to the trace on spins the kinetic exchange interaction is equal to $\frac{2t^2_{ij}}{\widetilde U}$.   
the second ferromagnetic term originates from the direct overlap of the neighboring Wannier functions.\cite{Mazurenko} One should note that due to the trace on spins the kinetic exchange interaction is equal to $\frac{2t^2_{ij}}{\widetilde U}$. 

In Table \ref{tab:Jresults}, we show the isotropic exchange interactions between the Wannier functions in C$_2$F and C$_2$H. 
 In agreement with Ref.~\onlinecite{Rudenko}, the resulting isotropic model for C$_2$F corresponds to the Heisenberg model on the triangular lattice with the nearest neighbor interactions. The kinetic contribution [the first term in Eq.~(\ref{exch})] to the total isotropic exchange interaction amounts to 40 meV, which is in excellent agreement with that presented in Ref.~\onlinecite{Rudenko} (note the difference in the spin Hamiltonian definition). 
However, we find that the kinetic antiferromagnetic coupling can be partially or fully compensated by the direct ferromagnetic exchange. Depending on the value of $J^{F}_{ij}$ the leading exchange interaction in C$_2$F between the nearest neighbors can be either antiferromagnetic or ferromagnetic.
In this situation, other types of magnetic couplings, for instance, the anisotropic (relativistic) exchange interaction can play an important role in formation of the magnetic structures in C$_2$F. 

In turn, the isotropic interaction for the 0-1 bond in C$_2$H is ferromagnetic, since the corresponding hopping integral is much smaller than that in C$_2$F.  The absolute value of $J_{01}$ is larger compared to the long-range antiferromagnetic couplings within the second and third coordination spheres.
Thus, the resulting spin model for C$_2$H is the Heisenberg Hamiltonian with the ferromagnetic nearest-neighbor, antiferromagnetic second- and third-neighbor interactions on the triangular lattice. Depending on the ratio between isotropic exchange interactions and the value of external magnetic field solutions of this model can reveal incommensurate spiral structures and skyrmion lattice states.\cite{Okubo}  

\begin{table}[h]
\caption{\label{tab:Jresults} Isotropic  exchange interactions (in meV) between the Wannier functions calculated by means of Eq.~(\ref{exch}) with the fully screened (bare) inter-site exchange interaction $J_{01}^F$.}
\begin{tabular}{ccc}
\hline
bond & $J_{ij}$ (C$_{2}$F) & $J_{ij}$ (C$_{2}$H)  \\
\hline
0-1  & 22 (-4)        & -33 (-98)       \\
0-2  &  0.020 &  7.26   \\
0-3  &  0.024 &  5.00\\
0-4 &  0.044  &  0.33\\
0-5 & 0.042  &   0.06 \\
\hline
\end{tabular}
\end{table}

{\it Dzyaloshinskii-Moriya interaction.}
Anisotropic exchange parameters can be derived by extending the theory of superexchange interaction in the case of spin-orbit coupling. They have the following form:\cite{Moriya, Aharony}
\begin{equation}
\label{DMI}
\boldsymbol{D}_{ij}=-\frac{i}{2 \widetilde U} [{\rm Tr_{\sigma}} \{ \hat t_{ji} \} {\rm Tr_{\sigma}} \{ \hat t_{ij}  \boldsymbol{\sigma} \}  -  {\rm Tr_{\sigma}} \{ \hat t_{ij} \} {\rm Tr_{\sigma}} \{ \hat t_{ji} \boldsymbol{\sigma} \}  ], 
\end{equation}
where $\boldsymbol{\sigma}$ are the Pauli matrices. Such an approach gives reliable results for the low-dimensional copper oxides.\cite{Nikolaev, Danis}
For the nearest neighbour bonds in C$_2$F with the radius vectors $\boldsymbol{R}_{01}$ = (1, 0, 0), $\boldsymbol{R}_{01'}$ = ($\frac{1}{2}$, -$\frac{\sqrt{3}}{2}$,0) and $\boldsymbol{R}_{01''}$ = ($\frac{1}{2}$, $\frac{\sqrt{3}}{2}$,0) we obtain $\boldsymbol{D}_{01}$ = (0, -0.93, -0.28), $\boldsymbol{D}_{01'}$ = (-0.81, -0.47, 0.28) and $\boldsymbol{D}_{01''}$ = (0.81, -0.47, 0.28) meV, respectively.  In contrast to C$_2$F, the magnitude of the Dzyaloshoinskii-Moriya interaction in the semihydrogenated graphene is much smaller, $\boldsymbol{D}_{01}$ = (0, -0.017, 0), $\boldsymbol{D}_{01'}$ = (-0.015, -0.008, 0) and $\boldsymbol{D}_{01''}$ = (0.015, -0.008, 0) meV. 

Higher orders in spin-orbit coupling, such as the symmetric anisotropic exchange interaction \cite{Moriya}, $\hat {\boldsymbol{S}}_i \overset{\leftrightarrow}{\Gamma}_{ij} \hat {\boldsymbol{S}}_j $ are small. Their magnitude is about 10$^{-3}$ meV (C$_{2}$F) and 10$^{-5}$ meV (C$_2$H), and thus the Dzyaloshinskii-Moirya interaction is the main source of the magnetic anisotropy in semifluorinated and semihydrogenated graphene.

Symmetry of the spin Hamiltonian is consistent with the $C_{3v}$ point group of the C$_{2}$F and C$_{2}$H systems (given by the vertical reflection planes going through the nearest-neighbor functionalized and non-functionalized carbon atoms and by $C_{3}$ rotations around adatom-carbon bonds). However, the resulting Wannier functions reside on the non-functionalized carbon sites and form a triangular lattice without reflection of the original lattice (Fig. \ref{ris:wannier_functions}). According to Moriya's rules\cite{Moriya}, since the reflection planes pass through the middle of the bonds between two Wannier functions, the corresponding  anisotropic exchange parameters lie in the reflection planes and are perpendicular to their bonds. Their directions are given arbitrarily with respect to the mirror planes, and as a result $z$ components of the anisotropic exchange parameters can alternate within the coordination sphere under $C_{3}$ rotations. Thus, the resulting symmetry of the effective model is $C_{3}$.

The electronic Hamiltonians constructed in the Wannier function basis also give us an opportunity to compare the spin-orbit coupling strength in the C$_2$H and C$_{2}$F systems. The previous first-principles studies \cite{Fabian1,Fabian2} have demonstrated an enhancement of spin-orbit coupling in C$_2$F compared to C$_2$H. Here, we confirm this finding by calculating the hopping integrals with spin-orbit coupling and estimating the Dzyaloshinskii-Moriya interactions.

\begin{figure}[t] 
	\includegraphics[width=0.9\linewidth]{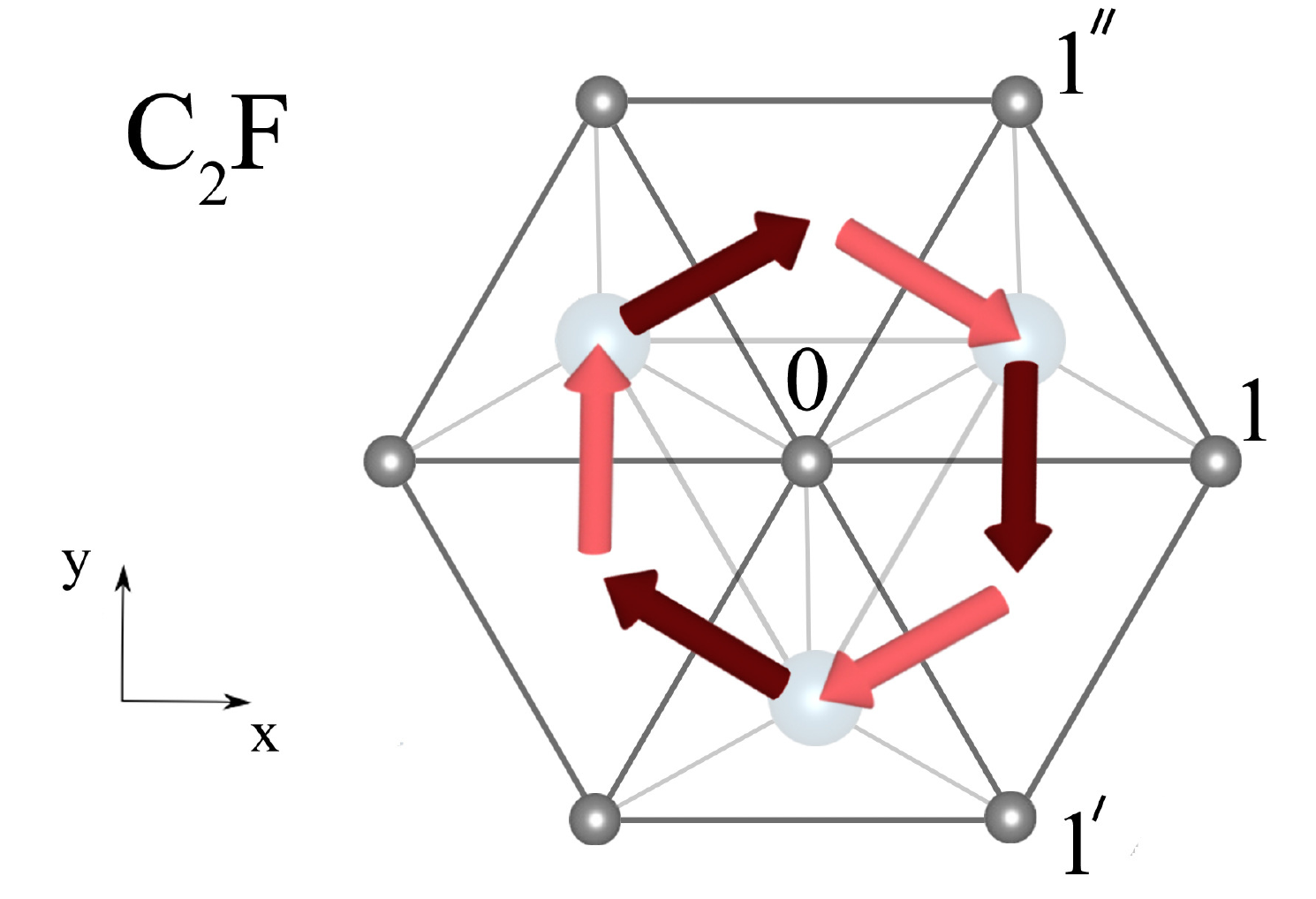}
	\caption{(Color online) Schematic representation of the Dzyaloshinskii-Moriya vectors in C${_2}$F. Light and dark red arrows denote the Dzyaloshinskii-Moriya vectors with positive and negative $z$ components, respectively.  The corresponding directions of the bonds are denoted with black arrows.}
	\label{DM_C2F} 
\end{figure}

We would like to stress that there is a conceptual difference in the origin of magnetic anisotropy in two-dimensional (2D) materials with $sp$ electrons that we consider and 2D materials with localized $d$ electrons. \cite{Wenzel,Torun}  While the magnetic anisotropy in $3d$, $4d$ and $5d$ systems originates from the spin-orbit coupling of individual metallic atoms, it is not the case in 2D materials with $sp$ electrons. As it was shown in our study there is a strong delocalization of the magnetic moments in C$_2$F and C$_2$H materials. Thus, the magnetic anisotropy in these $sp$ materials is a collective multi-atomic effect, which was also demonstrated by authors of Ref.\onlinecite{Fabian1, Fabian2} in their analysis of spin-orbit coupling in the C$_2$F and C$_2$H systems. In this situation the construction of the simple and transparent models for a system in question is a non-trivial task. By using the formalism of the Wannier functions we propose an elegant solution of the problem for C$_2$H and C$_2$F materials. The resulting electronic models for C$_2$H and C$_2$F are one-band Hamiltonians with spin-orbit coupling.

{\it Estimation of the $g$-factor}. 
To characterize orbital magnetism induced by fluorine atoms we have estimated the value of the $g$-factor. Moriya \cite{Moriya} used the gyromagnetic ratio that can be found from magnetic experiments to estimate a magnitude of the intersite anisotropic exchange interaction. In our case we are to solve an inverse problem. Having calculated the Dzyaloshinskii-Moriya interaction we can estimate the $g$-factor value for future magnetic experiments on C$_2$F. For that the quantity of interest is $\frac{|\boldsymbol{D}_{01}|}{J^{kin}_{01}}$ which is proportional to $\frac{\Delta g}{g}$, where $\Delta g$ is the deviation of the $g$-factor from the value for a free electron. For the semifluorinated graphene the estimated value of $g$ is about 2.025. As we will show below this information is important for estimating critical magnetic fields at which a skyrmion state is formed. 

\section{Hartree-Fock simulations}
To solve the electronic models given by Eq.~(\ref{Ham}) at zero temperature, we have employed the mean-field Hartree-Fock approximation:
\begin{equation}
\left(\hat{t}_{\boldsymbol{k}}+\hat{\mathcal{V}}^{H}_{\boldsymbol{k}} \right)|\varphi_{\boldsymbol{k}}\rangle=\varepsilon_{\boldsymbol{k}}|\varphi_{\boldsymbol{k}}\rangle,
\label{hf}
\end{equation}
\noindent where $\hat{t}_{\boldsymbol{k}}$ and $\hat{\mathcal{V}}^{H}_{\boldsymbol{k}}$ are the Fourier transforms of the hopping parameters and Hartree-Fock potential, respectively. $\varepsilon_{\boldsymbol{k}}$ and $|\varphi_{\boldsymbol{k}}\rangle$ are the corresponding eigenvalues and eigenvectors in the Wannier function basis; Eq.~(\ref{hf}) is solved self-consistently with respect to the density matrix:
\begin{equation}
\hat{n}=\sum\limits_{\boldsymbol{k}}|\varphi_{\boldsymbol{k}}\rangle \langle\varphi_{\boldsymbol{k}}|,
\end{equation}
\noindent and the resulting magnetic state is further defined as $\boldsymbol{S}=\mathrm{Tr}\left\{\hat{n}\boldsymbol{\sigma}\right\}/2$.
To find more detail on the computation scheme, we refer the reader to Refs.~\onlinecite{SolovyevHF, Nikolaev}. 

\begin{figure}[t] 
	\includegraphics[width=0.9\linewidth]{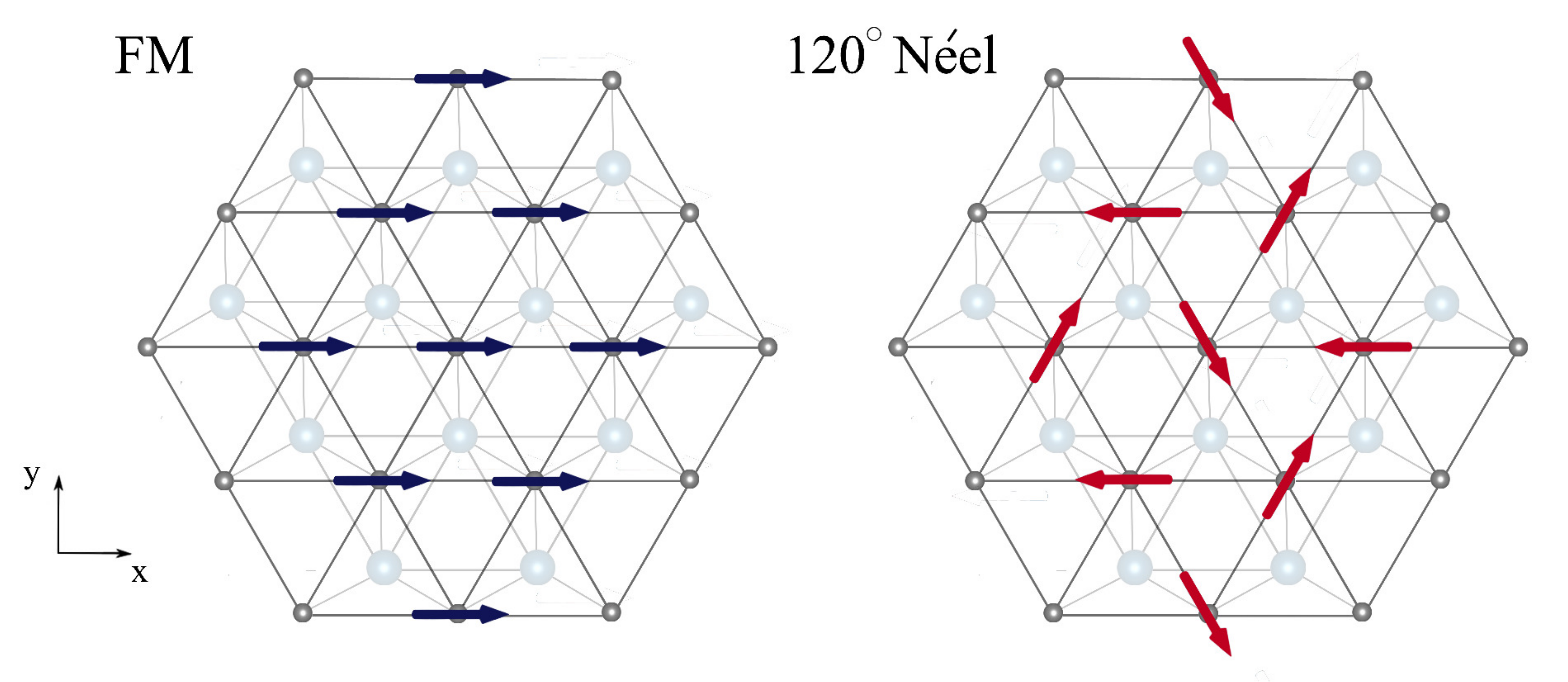}
	\caption{(Color online) The ferromagnetic (left) and  120$^{\circ}$ N\'eel (right) solutions obtained within the Hartree-Fock approximation.}
	\label{magn_config} 
\end{figure}

The results obtained for the semihydrogenated graphene with unit cells of different size do not reveal any signature of the spiral spin ordering, and the ferromagnetic state is stabilized [Fig.~\ref{magn_config}(left)]. The energy of the system does not depend on the direction of the total magnetization. From the analysis of the hopping integrals we conclude that spin-orbit coupling in C$_2$H is weak and does not produce magnetic anisotropy. Since the role of magnetic anisotropy is decisive in the formation of a long-range magnetic order in 2D materials \cite{Irkhin1,Irkhin2}, no long-range ordering is expected in C$_2$H at finite temperatures according to the Mermin-Wagner theorem \cite{Mermin}.

In the case of the semifluorinated graphene the situation is different. Depending on the choice of the direct exchange interaction $J^{F}_{01}$, one obtains either the 120$^{\circ}$ N\'eel for $J^{F}_{01} <$ 40 meV [Fig.~\ref{magn_config}(right)] or ferromagnetic states for $J^{F}_{01} >$ 40 meV [Fig.~\ref{magn_config}(left)]. There is a solution with zero isotropic magnetic interactions at $J^{F}_{01}$ = 40 meV, when the ferromagnetic direct exchange interaction exactly compensates the antiferromagnetic Anderson's superexchange. Thus we conclude that C$_2$F can be considered to be at the threshold of the antiferromagnetic-ferromagnetic instability.

\begin{figure*}[] 	
\includegraphics[width=1.0\linewidth]{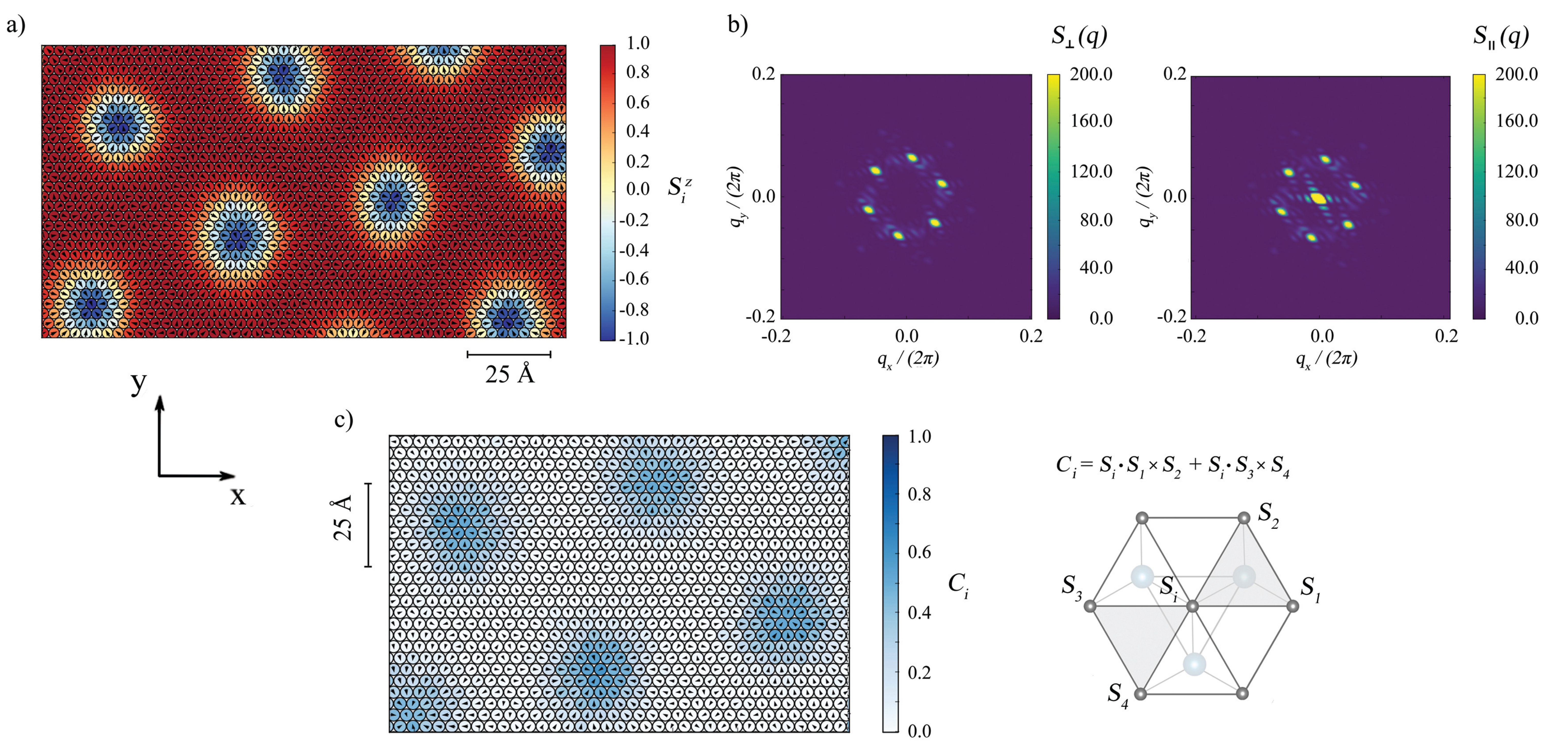}
	\caption{(Color online) Spin configuration (a), static spin structure factors (b) and local chirality (c) of the skyrmion phase realized in C$_{2}$F. Spin components in the $xy$ plane are indicated with black arrows.}
    \label{MonteCarloSki}
\end{figure*}

\section{Magnetic state at finite temperature and magnetic fields}
Our investigation of the semifluorinated and semihydrogenated graphene revealed that these systems are physical realizations of different spin models on the triangular lattice. Such a lattice is of special interest due to the effects of magnetic frustration and possibility to form topologically protected spin textures, skyrmions at finite temperatures and magnetic fields.  For instance, it was recently shown that skyrmionic states can be stabilized in the $J_1-J_3$ model with the ferromagnetic nearest neighbor and antiferromagnetic  next-nearest neighbour exchange interactions.\cite{Okubo} A similar model is derived in our study for C$_2$H. However, the ratio $\frac{|J_1|}{J_3}$ is much larger than that proposed in Ref.~\onlinecite{Okubo} to stabilize a skyrmion state. 

Another important example known from the literature is the antiferromagnetic (or ferromagnetic) triangular lattice with Dzyaloshinskii-Moriya interactions that favour formation of the antiferromagnetic skyrmion lattice state\cite{Rosales} (or a skyrmion lattice state with the N\'eel-type domain wall alignment \cite{Loidl}). A similar scenario can be realized in the C$_2$F system.  

To check whether it is possible to stabilize a distinct spin texture in the semifluorinated graphene we have performed classical Monte Carlo simulations based on the single-spin Metropolis update scheme and the heat bath method combined with overrelaxation for the spin models obtained with different values of $J^{F}_{01}$. In our calculations supercells of various size from $N=96\times96$  to $192\times192$ spins with periodic boundary conditions were used, and a single run contained $(0.2-1.0)\times 10^{6}$ Monte-Carlo steps.  While different states can be identified from a real space spin configuration, we have also computed the static spin structure factors:
\begin{equation}
S_{\perp}(\boldsymbol{q})=\frac{1}{N}\left \langle \left| \sum_{i}S_{i}^{x}e^{-i\boldsymbol{q}\cdot\boldsymbol{r}}\right|^{2}+\left| \sum_{i}S_{i}^{y}e^{-i\boldsymbol{q}\cdot\boldsymbol{r}}\right|^{2}\right\rangle
\end{equation}
\noindent and
\begin{equation}
S_{\parallel}(\boldsymbol{q})=\frac{1}{N}\left \langle \left| \sum_{i}S_{i}^{z}e^{-i\boldsymbol{q}\cdot\boldsymbol{r}}\right|^{2}\right\rangle,
\end{equation} where $\langle ... \rangle$  means the Monte-Carlo averaged configuration, as well as the so-called local chirality $C_{i}=\boldsymbol{S}_{i}\cdot \boldsymbol{S}_{1}\times \boldsymbol{S}_{3}+\boldsymbol{S}_{i}\cdot \boldsymbol{S}_{3}\times \boldsymbol{S}_{4}$ that is regarded as an order parameter of the corresponding magnetic state.

In the case of the C$_2$F spin models obtained with $J^{F}_{01} <$ 40 meV, the antiferromagnetic skyrmion lattice state is destroyed by a weak $z$ component of the Dzyaloshinskii-Moriya interaction, instead the 120$^{\circ}$ N\'eel state is observed.

The situation is different for the C$_2$F spin models with $J^{F}_{01} >$ 40 meV. In this case a N\'eel type skyrmion state can be realized. An example shown in Fig.~\ref{MonteCarloSki}(a) was obtained with the exchange interactions $J_{01}$ = 1.9 meV and $\boldsymbol{D}_{01}$ = (0, -0.93, -0.28) at the temperature $\frac{T}{|J_{01}|}$= 0.02 and the magnetic field $\frac{B}{|J_{01}|}$ = 0.1. Fig.\ref{MonteCarloSki}(b) gives the intensity of the spin structure factor for the obtained texture. There is a superposition of three spirals with  $\pm \mathbf{q}$ pairs of the wave vectors, which is a clear indication of the skyrmion lattice state.   

Taking into account the estimated value of the $g$-factor $g$=2.025, the critical magnetic field needed to stabilize the skyrmion lattice can be defined as 1.62 T. For this set of parameters the size of the individual skyrmion can be estimated to be about 25 \AA.
Generally, it is controlled by the ratio $\frac{\boldsymbol{D}_{01}}{J_{01}}$ and the magnetic field. 

\section{Conclusions}
Our theoretical results indicate that the recent experimental realization of C$_2$F \cite{Kashtiban} opens a way for exploiting truly two-dimensional one-band model materials demonstrating a rich variety of physical properties, such as the strong Dzyaloshinskii-Moriya interaction, spatial charge correlations, magnetic frustration, skyrmion state and others.   

Specifically, we found that the Wannier functions describing magnetic moments in the C$_2$F and C$_2$H systems form a triangular lattice with completely different hopping paths. The overlap of the neighbouring Wannier orbitals results in a strong direct ferromagnetic exchange interaction for both systems in question that can partially or fully compensate the kinetic superexchange interaction between nearest magnetic orbitals. For the semihydrogenated graphene we predict a short-range ferromagnetic order.
Since our Hartree-Fock calculations do not reveal any magnetic anisotropy in C$_2$H, the long-range magnetic order in this system is unstable as follows from the Mermin-Wagner theorem \cite{Mermin} for two-dimensional systems.   

The calculations of the anisotropic exchange interactions suggest that C$_2$F can be considered as a physical realization of the Heisenberg model with DMI on the triangular lattice. The variation of $J^{F}_{ij}$ from the fully screened to bare limits leads to either the 120$^{\circ}$ N\'eel or skyrmion states as it was predicted by our Hartree-Fock and Monte Carlo simulations.

The key parameter that is responsible for a variety of magnetic states in C$_2$F and C$_2$H is the direct exchange interaction, $J^F_{ij}$ between the neighbouring Wannier functions. Experimentally, one can find different mechanisms to control and tune this interaction. For instance, in this paper we demonstrate that $J^F_{ij}$ is sensitive to the screening by the background. The latter can be changed by coupling with a substrate. \cite{Rudenko1} 
Another attractive control parameter is strain that strongly affects non-local Coulomb interactions in graphene systems as it was shown in Ref.~\onlinecite{Wehling}.  

\begin{acknowledgments}
We acknowledge fruitful communications with Igor Solovyev. The hospitality of the Institute of Theoretical Physics of Hamburg University and Radboud University of Nijmegen is gratefully acknowledged. The work is supported by the Ministry of Education and Science of the Russian Federation, Project No. 16.1751.2014/K and the grant of the President of Russian Federation MD-6458.2016.2. A.I.L. acknowledges the support of DFG Priority Programme 1459. A.N.R. and M.I.K. acknowledge support from the European Union, Horizon 2020 research and innovation programme under grant agreement No. 696656, GrapheneCore1.
\end{acknowledgments}

\end{document}